\documentclass[11pt]{article}
\newcommand{\be}{\begin{equation}}
\newcommand{\ee}{\end{equation}}
\newcommand{\beq}{\begin{eqnarray}}
\newcommand{\eeq}{\end{eqnarray}}

\def\theequation{\arabic{section}.\arabic{equation}}
\usepackage{graphicx}

\begin{document}
\begin{center}
{\bf \LARGE Summing all the eikonal graphs. II}\\
[30mm]
H. M. FRIED\\
{\em Physics Department \\
Brown University \\
Providence R.I. 02912 USA}\\
[5mm]
Y. GABELLINI\\
{\em Institut Non Lin\'eaire de Nice\\
 1361 Route des Lucioles\\
06560 Valbonne France}\\
[25mm]
Abstract
\end{center}

	A physically reasonable model is introduced in order to estimate, in a functional way, the vast number of distinct graphs which are conventionally neglected in eikonal scatttering models that lead to total cross sections increasing with energy in the form of the Froissart bound. A range of estimates of non leading--log effects on the conventional, leading--log, ladder--tower graphs is also attempted. Upon summing $\underline {\rm all}$ the eikonal graphs of this model, one finds significant cancellations away from the conventional, tower graph result, with total cross sections tending to constant values at extreme scattering energies on the order of $10^3 - 10^6\  TeV$

\newpage

{\bf\section{Introduction}}
\setcounter{equation}{0}

It was pointed out some three decades ago \cite{one} that the currently and continually popular  `` tower graph '' eikonal scattering model \cite{two} contains within it a pair of fundamental omissions, which can be simply characterized as (i) the absence of all but leading--logarithm (LL) graphs (that is, only the leading $\ln(s/m^2)$ dependence of each perturbative,  `` ladder graph~'' order is retained, where $s$ is the total center--of--mass energy square, and $m$ is a convenient scale setting mass); and (ii) all of the contributions of the many, many other, non tower graphs are neglected. These approximations are understandable, in view of the immense complexity involved in extracting the sum of high order perturbative contributions; and it has been most fortunate that the tower graphs so approximated have provided a theoretical framework \cite{three} for understanding essentially all experimental hadronic scattering at energies above a few $GeV$.

Nevertheless, approximations (i) and (ii) have long represented open questions, whose answers have long been believed to be obtainable only in a functional approach, rather than by summing the contributions of individual, perturbative graphs. Precise answers were not possible at the time of the work of ref.\cite{one}, which was only able to point out the wide variety of possible forms for the scattering amplitude -- and, in particular, for the predicted total cross section, which is the physical quantity of interest emphasized here -- and for which it was shown that a wide variety of answers were possible, many of them differing from the Froissart bound of the tower graphs. The model analysis used in \cite{one} was part functional and part perturbative, and we here consider it to be the first paper of this `` series '', with the present paper, designated II, as the continuation of that work, some three decades later.

The reason that such a continuation is now possible is due to a recent paper by Tomaras, Tsamis and Woodard \cite{four}, generalizing the exact, half century old, Schwinger calculation of the probability of $e^+e^-$ production in the presence of a  constant electric field, to the situation where the electric field can depend upon either of the light-cone coordinates, $x_+$ or $x_-$ (but not both). This latter calculation was then repeated using an independent functional approach \cite{five}; and it was subsequently realized that the solubility contained in the functional light-cone model is sufficient to provide a decent representation of inelastic particle production in a mutiperipheral manner. By `` decent '' one means that the essential Physics should be preserved, even though relevant cut--offs may be necessary; and for the first time it now becomes possible, in this context, to carry through a complete functional estimation of the sum of $\underline {\rm all}$ the eikonal graphs.

The result of these calculations suggests, very strongly, that `` internal, unitarity '' cancellations will, with rising $s$, eventually become so strong that all $\sigma_{\scriptscriptstyle TOT}(s)$ will start to fall off and tend to constant values. Just where this happens depends upon how closely the present, `` scalar pion '' model approximates full QCD; but if this model calculation is matched in a reasonable way to experimental data, one sees that the fall off cannot begin until well past the $s$ value of the so--called `` cosmic ray point ''. Hence, at present, this effect would not be directly measurable, although it may have some  relevance to cosmic ray cross sections at extremely high energies. Nevertheless, any qualitative, or even cruder answer to the questions raised by the approximations (i) and (ii) above is not without interest.

\vskip1cm   
{\bf\section{Formulation}}
\setcounter{equation}{0}
A brief but self--contained, functional derivation of eikonal scattering amplitudes has been given elsewhere \cite{seven},  and need not be repeated here; this subject requires a certain familiarity with the $S$ Matrix, and the way in which its elements may be expressed in terms of appropriate, mass--shell amputated, $n$--point functions of Quantum Field Theory. Perhaps the most comprehensive collection of eikonal references may be found in the book by Cheng and Wu \cite{eight}, whose later chapters describe the perturbative calculations which have been made for the eikonal function in QCD. The intent of this section's remarks is to set the stage for an explicit, functional representation of the eikonal for high energy scattering in the so--called `` multiperipheral model~'', with calculations appearing in the subsequent sections.

Aside from multiplicative and renormalization constants, the essence of the connection between Green's functions and corresponding $S$ Matrix elements, in the limit of large scattering energy and small momentum transfer, lies in the exponential factors appearing in the mass--shell amputated Green's function for each (in this case) fermion entering and leaving the scattering region. When the scattering fermion couples to a Neutral Vector Meson (NVM), and the exchange of arbitrary numbers of NVMs between a pair of scattering fermions is desired, eikonalization of the amplitude takes place such that the scattering amplitude is expressed in terms of an eikonal function
\be
T(s,t) = {is\over 2m^2}\int \!d^2b\,e\,^{\displaystyle i\vec q_{\perp}\!\!\cdot\! \vec b}\,\Bigl[1 - e\,^{\displaystyle i\chi(s,\vec b)}\Bigr]
\ee
where $-t = q^2$ and $s=$ (total center--of--mass energy)$^2$. The eikonal is a function of impact parameter $b$, and of $s$; and for the simple case of the exchange of an arbitrary number of virtual NVMs between the scattering fermions, one finds the result
\be
i\chi_1(s,b) = -{ig^2\over 2\pi}\,\gamma(s) \,K_0(Mb),\ \ \ \ \ \  \gamma(s) = {(s - 2m^2)\over\sqrt{s(s-4m^2)}}
\ee
where $g$ is the fermion--NVM coupling constant, and $M$ denotes the NVM mass; to this eikonal there correspond the graphs of Fig.1. At high energies, the invariant differential cross section is given by 
$d\sigma/dt = (m^4/ \pi s^2)\vert T \vert^2$, while the total cross section is given by
\be
\sigma_{\scriptscriptstyle TOT}(s) = 2\, {\rm Re}\!\int \!d^2b\,\Bigl[1 - e\,^{\displaystyle i\chi(s,\vec b)}\Bigr]
\ee
Equations (2.1) and (2.3) are generic results, for any eikonal function, while (2.2) denotes the eikonal built from virtual NVM exchange between the scattering fermions.

\begin{figure}[htpb]
\centering\includegraphics[height=4cm,clip=true]{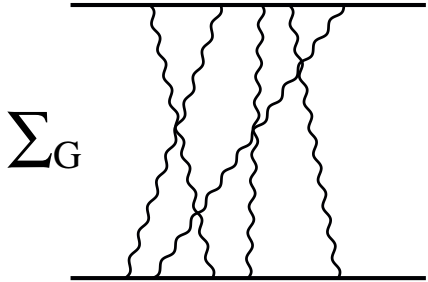}
\caption{Virtual NVMs exchanged between scattering particles}
\end{figure}

It should be noted that the total cross section calculated for the eikonal of (2.2) becomes a constant, independent of $s$ as $s/m^2\rightarrow\infty$, a fact which can be understood physically by the observation that the only inelastic graphs of this model are those of bremsstrahlung, and the latter always vanish for zero momentum transfer.

The important observation made by Cheng and Wu was that, in massive NVM QED, there is another type of inelastic process which can contribute to inelastic production at small $\vert t\vert/s$, the so--called `` multiperipheral '' graphs pictured in Fig.2a, which, by unitarity, correspond to the `` inelastic shadow~'' graphs of Fig.2b, graphs that must diminish the elastic amplitude if such inelastic production increases as energies increase. One there finds a phase space factor proportional to $\ln(s/m^2)$ for the probability of each fermion pair to be produced in this way, which suggests that the graphs of Fig.2b are the relevant graphs calculated by Cheng and Wu in massive--photon QED, and yielded an eikonal of form
\be
i\chi_2(s,b) = -a\,s^{\alpha} \,e\,^{\displaystyle -\mu b}
\ee
with $a$, $\alpha$, $\mu$ constants, which leads directly (Appendix A) to the estimate that $\sigma_{\scriptscriptstyle TOT}(s)\sim \ln^2s + \cdots$ in the limit of very high energies \cite{eight}.

\begin{figure}[htpb]
\centering\includegraphics[height=4cm,clip=true]{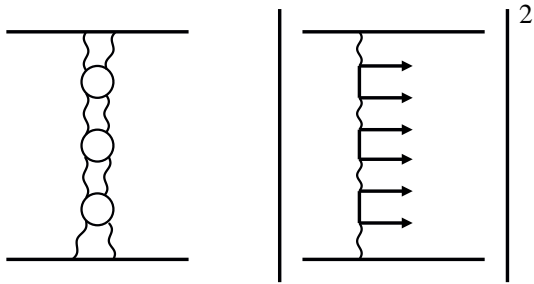}
\caption{a)(left) Tower of closed fermion loops. b)(right) Absorptive part of fermion-loop tower.}
\end{figure}

Shortly after this observation, it was pointed out by various authors \cite{nine} that another, and simpler form of  multiperipheral interaction generates a very similar result; this appears if scalar particles are exchanged between the NVM pairs, which are themselves exchanged between the scattering fermions, as in Fig.3a. Again, the unitarity `` shadow '' corresponds to inelastic graphs of the form shown in Fig.3b, where the phase space of each scalar particle emitted contributes a factor of $\ln(s/m^2)$. The corresponding eikonal, constructed from the graphs of Fig.3b takes the form
\be
i\chi_2\simeq - {a\over \ln(s/s_0)}{\Bigl({s\over s_0}\Bigr)}^{\alpha}\,e\,^{\displaystyle -\mu^2 b^2/\ln(s/s_0)}
\ee
with $a$, $\alpha$, $\mu$, $s_0$ constants, and leads to a very similar elastic amplitude, and to the same form of high energy $\sigma_{\scriptscriptstyle TOT}(s)$ as that of Cheng and Wu.

\begin{figure}[htpb]
\centering\includegraphics[height=4cm,clip=true]{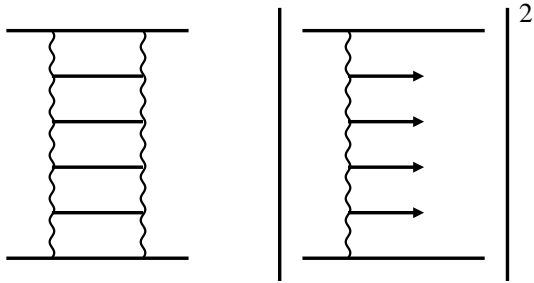}
\caption{a)(left) Tower of scalar particles exchanged between NVMs. b)(right) Absorptive part of this tower.}
\end{figure}

A functional description of all the eikonal graphs, including those not considered by Cheng and Wu and successors, has been given in references \cite{one} and \cite{seven}, and can be expressed in simple functional language. One begins by considering fermions coupled to NVM fields, and the latter coupled to ``  scalar pion '' fields; and one then organizes the functional pieces that are involved in the desired scattering amplitude, discarding all terms which correspond to more structure than that of NVMs exchanged between a pair of scattering fermions, with all possible virtual $\pi$ exchanges between all possible NVMs. In analogy with the forms quoted above, the NVM interactions eikonalize, carrying with them the composite substructures corresponding to multiple $\pi$ exchange between all possible NVMs, with a resulting expression for the eikonal 
\be
e\,^{\displaystyle i\chi} = \exp\biggl[-{i\over 2}\int \!{\delta\over\delta\pi}D_c{\delta\over\delta\pi}\biggr]\,\exp\biggl[i\int \!f_I\,{\overline\Delta_c}[\pi]\,f_{II}\biggr]\biggl\vert_{\pi\rightarrow 0}
\ee
where ${\overline\Delta_c}(x,y\,\vert \pi)$ denotes the propagator of a NVM of mass $M$ in the presence of a (fictitious) field $\pi(z)$, $D_c$ is the free propagator of a scalar pion of mass $\mu$, and the $f^{\mu}_{I,II}(z)$ denote the classical currents of the two fermions.

In this eikonal context, (2.6) is an exact result. The only approximation to (2.6) that will be considered here -- and only temporarily -- will be to discard all terms which correspond to self--energy structure of virtual $\pi$ emission an absorption along each NVM. The reason for this approximation is that we wish to begin with an estimation as close as possible to the LL ladder graphs of reference \cite{nine}, in which the only graphs considered correspond to virtual $\pi$ exchanges between a pair of NVMs, and in all possible ways (ladder plus all possible crossed graphs). But then, attempting to calculate this eikonal generalization of (2.5) -- which is given by expanding the right hand side of (2.6) to its quadratic dependence on ${\overline\Delta_c}[\pi]$ -- we shall find it most appropriate to return to the more exact form which retains self linkages along each NVM propagator. Finally, when the generalization to $\underline {\rm all}$ the eikonal graphs is attempted, we shall work with the complete eq.(2.6), modeling self linkages along with all possible $\pi$ exchanges, between all possible numbers of NVMs.

The inadequacy of previous LL approximation ladder graph calculations, which have made use of the unjustified approximation of retaining only the ``~leading--log~'' terms of every perturbative order, may be pointed out in the following way. If the coupling constant of the NVM to the scalar particle is $G$, a ladder graph with $n$ rungs contributes an amount proportional to $G\,^{2n}[\ln(s/m^2)]^n$, in contrast to a ladder in which one pair of rungs is crossed, of contribution proportional to $G\,^{2n}[\ln(s/m^2)]^{n-1}$; every time another pair of rungs is crossed, the $\ln(s)$ dependence drops by another power. The total number of such graphs is given by $n!$, and if only those terms with the largest powers of $\ln(s/m^2)$ are kept, this means that $n! - 1$ ``~less important~'' terms are discarded, an approximation that is mathematically untenable for $n$  sufficiently large such that $n! - 1$ is on the order of $\ln^{n-1}(s)$, or larger. Nevertheless, for reasons of `` simplicity '' -- one calculates what one can, and hopes for the best -- this type of approximation has long been made, without justification.

The second approximation mentioned in Section 1, made for the same reason of calculational simplicity, has been to neglect contributions coming from eikonal graphs more complicated than the towers. To even attempt such a calculation one is forced into a functional description, for the number of classes of graphs which must be included, corresponding to the exchange between scattering fermions of all possible $t$ channel NVMs, between which are exchanged all possible numbers of scalar mesons, is simply staggering. Such a functional description, given in (2.6), has been known for three decades; but what was lacking was a suitable NVM Green's function ${\overline\Delta_c}(x,y\,\vert \pi)$ corresponding to NVM propagation in a fictitious scalar field $\pi(x)$, which could be used to model the inelastic production of scalar particle, here called ``~scalar pions '', in order to estimate the elastic scattering eikonal of (2.6). Such a Green's function is presented immediately below, and is used to suggest one possible form of the eikonal in a `` generalized '' Cheng--Wu context, containing towers formed from ladders and crossed rung ladders, in all possible combinations, and to compare the result with that of the leading--log ladder graph eikonals. In the next Section, the calculation is extended to include the sum of all eikonal graphs of this model, without exception.

As a preliminary step, we remind the reader of the functional cluster expansion discussed in detail in the second book of reference  \cite{seven}, in particular to 
\be
\exp\biggl({\displaystyle -{i\over 2}\int \!{\delta\over\delta\pi}D_c{\delta\over\delta\pi}}\biggr)\,\exp\bigl({\displaystyle L[\pi]}\bigr) = \exp\biggl({\displaystyle \sum_{n=1}^{\infty} Q_n/n!}\biggr)
\ee
where the $Q_n$ are the appropriate, connected, cluster functionals. In the present context, the functional $L[\pi]$ is given by the right hand side exponential factor of (2.6), and is operated upon by the linkage operator, as shown. At first, we will be interested only in the exchange of virtual $\pi$s between the virtual NVMs -- that is where the Physics of the LL ladder--tower graphs lies, governed by unitarity -- and we first drop all radiative corrections along each NVM, which is a simplification that can easily be performed functionally. In particular, the NVM mass $M$ and coupling $G$ to the pions are taken as ``~renormalized '' constants ( at least until the end of the next Section, when a closer correspondence of this model with QCD is attempted ). 

Dropping all radiative corrections along the NVMs, the quantity $Q_1$, as defined in (2.7), becomes just the $i\chi_1$ of (2.2), while the $Q_2$ of (2.7) may be written as
\be
Q_2 =\biggl[ e\,^{\displaystyle -i\int \!{\delta\over\delta\pi_a}D_c{\delta\over\delta\pi_b}} -1\biggr]L[\pi_a]L[\pi_b]\biggl\vert_{\pi_{a,b}=0}
\ee
and is the `` tower graph '' approximation to the eikonal of this problem, with all numbers of virtual $\pi$s exchanged, as ladders and crossed rung ladders, between one pair of NVMs. The $Q_n$ for higher powers of $n$ correspond to corrections to this tower eikonal, constructed from $\pi$--exchange between more than two NVMs, and in all possible ways  between these multiple NVMs.

The next step is the specification of a suitable Green's function, ${\overline\Delta_c}[\pi]$, which can model the emission of relatively high energy pions from each NVM. Experimentally, most of the momenta of particles emitted inelastically is in close--to--forward directions, and so we may imagine that the field $\pi(x)$ depends only on $x_3$ and $x_0$, with transverse momentum components subsequently limited ( which is also in agreement with experimental inelastic emissions ) in another, model--dependent way. Because these emitted particles ( gluonic jets, in QCD ) are of high energy -- or, more accurately, we wish to extract those parts of these individual processes which increase as $\ln(s)$ -- we can assume that these are all relativistic particles, and replace $\pi(x_3,x_0)$ by $\pi(x_3-x_0)$.

This form suggests particles moving relativistically in the $+x_3$ direction; but whether that direction lies in the $\pm z_3$ direction of the center--of--mass depends on how these $x$--variables connect to the pion propagator $D_c(u-v)$. The latter is perfectly relativistic, in the sense that it contains both particle and anti--particle poles in its $k_0$--plane; and it will generate a logarithmic divergence corresponding to particles emitted in the $\pm z_3$ directions, as dictated by energy--momentum considerations. That log divergence corresponds to one that would be found in the probability for emitting a scalar pion of arbitrarily high longitudinal momentum; and just as is done for the ordinary tower ( ladder ) graphs, we regulate that log divergence by the physically sensible requirement that each $(k_3)_{max}\sim \sqrt s$. The model is completed by inserting, by hand, a $k_{\perp}$ cut--off in the definition of $D_c$, and one can then study the effects of doing this for different types of $k_{\perp}$ cut--offs.

These physically--motivated restrictions and insertions define the model, which can then be used to reproduce the essential results of the ladder graph towers, and to explore the `` internal, unitarity cancellations '' which one might expect to result from summing over all the eikonal graphs, as in the next Section. However, even with the physical attributes of this model, one is not yet able to perform the full analysis without an additional assumption, which will shortly become clear.

The model's simplicity can now be realized by rewriting $\pi(x_3-x_0)$ as $\pi(n^{(-)}\!\cdot x)$, and by recognizing the similarity of this Green's function with that of the `` scalar laser '' $G_c[A]$, one of the very few Green's functions that can be obtained explicitly; here, $A(x)\rightarrow A(k\!\cdot \!x) = A(n^{(-)}\!\cdot \!x)$, for $k_{\mu} = (0,0,\omega;i\omega)$. All the results of that analysis, reproduced for completeness in Appendix B, may be taken over immediately by replacing $k_{\mu}$ of (B.7) by $k_{\mu}/\omega$ and $A(x)$ by $\pi(x)$, so that one can write ${\overline\Delta_{c,\mu\nu}} = \delta_{\mu\nu}\,{\overline\Delta_c}$, with
\be
{\overline\Delta_c(x,y\vert \pi)} = {1\over 16\pi^2}\int_0^\infty\!{ds_a\over s_a^2}\,e\,^{\displaystyle -is_am^2\!+\! i{(x-y)^2\over 4s_a}} e\,^{\displaystyle -iGs_a\!\int_0^1\! \!d\lambda\, \pi(n^{(-)}\!\cdot \!\xi(x,y\vert\lambda))}
\ee
where $\xi_{\mu} = \lambda\, x_{\mu} + (1 - \lambda)\,y_{\mu}$, and where we will use the subscripts $a, b, c\ldots$ to distinguish the different NVM propagators.
\vskip1cm   
{\bf\section{Tower--graph calculations}}
\setcounter{equation}{0}
Each such Green's function enters into (2.6) in the form 
\beq
{}&\displaystyle ig^2(p_1\!\cdot\!p_2)\int\!\!\!\int_{-\infty}^{+\infty}\!\!d\bar s\,d\bar t\,\delta(u-[z_1 - \bar sp_1])\delta(v-[z_2 - \bar tp_2])\overline\Delta_c(u,v\vert \pi)\nonumber\\
&=\displaystyle i{g^2(p_1\!\cdot\!p_2)\over 16\pi^2}\int_0^{\infty}{ds_a\over s_a^2}\,e\,^{\displaystyle -is_am^2}\times\nonumber\\
&\times\displaystyle\int\!\!\!\int_{-\infty}^{+\infty}\!\!d\bar s_a\,d\bar t_a\,e\,^{\displaystyle (i/4s_a)(z_{12}-\bar s_ap_1+\bar t_ap_2)^2}\,e\,^{\displaystyle -iGs_a\!\int_0^1\! \!d\lambda_a\, \pi(n^{(-)}\!\cdot \!\xi_a(\lambda_a))}\nonumber
\eeq
where $z_{12}=z_1-z_2$ and $\xi_a(\lambda_a)=\xi_a(z_1-\bar s_ap_1,z_2-\bar  t_ap_2\vert\lambda_a)$. Restricting the calculation to $Q_2$, which will yield this model's version of $i\chi_2$, from (2.8) we need calculate
\beq
{}&\displaystyle \biggl[i{g^2(p_1\!\cdot\!p_2)\over 16\pi^2}\biggr]^2\int_0^{\infty}{ds_a\over s_a^2}\int_0^{\infty}{ds_b\over s_b^2}\,e\,^{\displaystyle -i(s_a+s_b)m^2}\int\!\!\!\int_{-\infty}^{+\infty}\!\!d\bar s_a\,d\bar t_a\int\!\!\!\int_{-\infty}^{+\infty}\!\!d\bar s_b\,d\bar t_b\nonumber\\
&\times\displaystyle e\,^{\displaystyle (i/4s_a)(z_{12}-\bar s_ap_1+\bar t_ap_2)^2}e\,^{\displaystyle (i/4s_b)(z_{12}-\bar s_bp_1+\bar t_bp_2)^2}\biggl[ e\,^{\displaystyle -i\int \!\!{\delta\over\delta\pi_a}D_c{\delta\over\delta\pi_b}} -1\biggr]\nonumber\\
&\times\displaystyle e\,^{\displaystyle -iGs_a\!\int_0^1\! \!d\lambda_a\, \pi_a(n^{(-)}\!\cdot \!\xi_a(\lambda_a))}\,e\,^{\displaystyle -iGs_b\!\int_0^1\! \!d\lambda_b\, \pi_b(n^{(-)}\!\cdot \!\xi_b(\lambda_b))}\biggl\vert_{\pi_{a,b}=0}
\eeq
where $\tilde D_c(k)=e\,^{\displaystyle -\gamma^2 k_{\perp}^2}[\mu^2+k_3^2-k_0^2-i\epsilon]^{-1}$, and $\gamma$ is the $k_{\perp}$ cut--off to be specified below.

The linkage operator $\biggl[ e\,^{\displaystyle -i\int \!{\delta\over\delta\pi_a}D_c{\delta\over\delta\pi_b}} -1\biggr]$ acting on the two last exponentials of (3.1) gives
\be
\displaystyle e\,^{\displaystyle iG^2s_as_b\!\int_0^1\! \!d\lambda_a\int_0^1\! \!d\lambda_b\, D_c(\xi_a^{(-)}(\lambda_a)-\xi_b^{(-)}(\lambda_b))}-1
\ee
while the propagator of (3.2) may be written as
\be
\int{d^2k_{\perp}\over (2\pi)^2}\,e\,^{\displaystyle -\gamma^2 k_{\perp}^2}\,{1\over 2}\int\!\!\!\int{dk_{+}dk_{-}\over (2\pi)^2}{e\,^{\displaystyle (i/2)k_{+}(\xi_a^{(-)}(\lambda_a)-\xi_b^{(-)}(\lambda_b))}\over \mu^2+k_{+}k_{-}-i\epsilon}
\ee
where $k_{\pm}=k_3\pm k_0$. With ${\mathcal Z}={(\xi_a^{(-)}-\xi_b^{(-)})/2}$, the $k_{+}$ integrals  of (3.3) may be written as 
$${1\over 2}{1\over (2\pi)^2}\int{dk_{-}\over k_{-}}\int \!dk_{+}\, e\,^{\displaystyle ik_{+}{\mathcal Z}}\biggl[{\mu^2\over k_{-}}+k_{+}-i\epsilon\!\cdot\!\epsilon(k_{-})\biggr]^{-1}$$
where $\epsilon\rightarrow 0^{+}$, and $\epsilon(x)=\theta(x)-\theta(-x)$. Integration over $k_{+}$ depends on the sign of $k_{-}$ and yields
\be
(2i\pi)\biggl\{\theta(k_{-})\,e\,^{\displaystyle -i\vert{\mathcal Z}\vert\mu^2/k_{-}}-\theta(-k_{-})\,e\,^{\displaystyle +i\vert{\mathcal Z}\vert\mu^2/k_{-}}\biggr\}
\ee
so that both terms of (3.4) contribute equally to the remaining $k_{-}$ integral, yielding
\be
{i\over 2\pi}\int_0^{\infty}{dk\over k}\,e\,^{\displaystyle -i\vert{\mathcal Z}\vert\mu^2/k}
\ee
This integral diverges logarithmically for large $k$, and as explained above, we insert a cut--off $k_{max}\sim\sqrt s$, and an arbitrary scale parameter $m$, to obtain the dominant contribution for large $s$: $\displaystyle(i/4\pi)\ln\left(s/m^2\right)$. In this way, (3.3)  becomes $\displaystyle\left(i/(4\pi\gamma)^2\right)\ln\left(s/m^2\right)$; and because this leading $s$--dependence is independent of $\lambda_{a,b}$, (3.2) simplifies to 
\be
\exp\left[-\alpha_G\left({m^2s_as_b\over 4\pi\gamma^2}\right)\ln\left(s\over m^2\right)\right] - 1
\ee
where $\alpha_G = G^2/4\pi m^2$, and, subsequently, $\alpha_g = g^2/4\pi$.

The parametric integrals over $s_{a,b}$, $t_{a,b}$ still remain to be done, with (3.1) replaced by
\beq
{}&\displaystyle {-\alpha_g^2s^2\over 4(4\pi)^2}\int_0^{\infty}{ds_a\over s_a^2}\int_0^{\infty}{ds_b\over s_b^2}\,e\,^{\displaystyle -i(s_a+s_b)m^2}\biggl[e\,^{\displaystyle -\left({m^2\alpha_G\over 4\pi\gamma^2}\right)s_as_b\ln\left({s\over m^2}\right)} - 1\biggr]\nonumber\\
&\times\displaystyle\int\!\!\!\int_{-\infty}^{+\infty}\!\!d\bar s_a\,d\bar t_a\int\!\!\!\int_{-\infty}^{+\infty}\!\!d\bar s_b\,d\bar t_b
\displaystyle \exp\left[{\displaystyle {i\over 4s_a}(z_{12}-\bar s_ap_1+\bar t_ap_2)^2 + {i\over 4s_b}(z_{12}-\bar s_bp_1+\bar t_bp_2)^2}\right]\nonumber
\eeq
and with $p_{1\mu} = En_{\mu}^{(-)}$, $p_{2\mu} = -En_{\mu}^{(+)}$, $\vec b =\vec z_{12}$, those integrals display a lovely cancellation of all non transverse $z_{12}$ dependence, and generate 
\beq
&i\chi_2={\displaystyle - \alpha_g^2\int_0^{\infty}{ds_a\over s_a}\int_0^{\infty}{ds_b\over s_b}\,e\,^{\displaystyle -i(s_a+s_b)m^2 + i{b^2\over 4}\biggl({1\over s_a} + {1\over s_b}\biggr)}}\nonumber\\
&\times\biggl[e\,^{\displaystyle -\alpha_G\left({m^2s_as_b\over 4\pi\gamma^2}\right)\ln\left(s\over m^2\right)} - 1\biggr]\nonumber
\eeq
or
\beq
&i\chi_2={\displaystyle - \alpha_g^2\sum_{n=1}^{\infty}{1\over n!}\left({m^2\alpha_G\over 4\pi\gamma^2}\right)^n\ln^n\left(s\over m^2\right)}\nonumber\\
&\\
\noalign{\vskip-0.4cm}
&\times{\displaystyle\int_0^{\infty}{ds_a\over s_a}\int_0^{\infty}{ds_b\over s_b}(-s_as_b)^n\,e\,^{\displaystyle -i(s_a+s_b)m^2}\,e\,^{\displaystyle i{b^2\over 4}\biggl({1\over s_a} + {1\over s_b}\biggr)}}\nonumber
\eeq
It is now convenient to make the standard continuation: $s_a\to -i\tau_a$, $s_b\to -i\tau_b$, so that (3.7) becomes
\be
\hskip-.3truecm i\chi_2={\displaystyle - \alpha_g^2\sum_{n=1}^{\infty}{1\over n!}\left({m^2\alpha_G\ln\left(s/ m^2\right)\over 4\pi\gamma^2}\right)^n\biggl[\int_0^{\infty}{d\tau\over \tau}\,\tau^n\,e\,^{\displaystyle -m^2\tau - b^2/4\tau}\biggr]^2} 
\ee
Since the integral inside the squared bracket of (3.8) is proportional to the Bessel function $K_n(mb)$, an alternate expression is
\be
i\chi_2={\displaystyle - 4\alpha_g^2\sum_{n=1}^{\infty}{1\over n!}\left({\alpha_G\over 16\pi}\biggl[{b^2\over\gamma^2}\biggr]\ln(s/m^2)\right)^nK_n^2(mb)}
\ee
This eikonal is properly absorptive, but -- were the sum over all $n$ performed ( or retained ) at the outset -- the final integrals over the $\tau$ variables appear to diverge in the large $\tau$ regions. But since we expect a strong correlation between the behavior of $\sigma_{\scriptscriptstyle TOT}(s)$ and large $b$ values $( b\sim b_0(s)\sim \ln(s/m^2))$, we can ask if (3.9) simplifies in the limit of large $b$; this would be the case for $b\gg 1/m$, which is certainly expected, since one assumes that $\ln(s/m^2)\gg1$. Note that, for large impact parameter, the only natural $k_{\perp}$ cut--off in the model propagator is $b$ itself, and this is our choice: $\gamma = b$.

One would then like to be able to replace each $K_n(mb)$ of (3.9) by its large  $mb$ asymptotic form $\displaystyle\left[{\pi/(2mb)}\right]^{1/2}\!e\,^{\displaystyle -mb}\,[1+\cdots]$, but any such interchange of sum and asymptotic limit must be viewed with suspicion, and justified, even if the result is so reasonable that it is not difficult to suppress disbelief. The mathematically improper step that one would like to take, for $mb\gg1$, is to replace each $K_n(mb)$ of (3.9) by its asymptotic form above, with leading term independent of $n$, so that the sum again exponentiates. This is correct only if $mb > (n^2 - 1)/2$, as is easily seen by examining the next terms of the expansion \cite{ten}. Were there only a finite number of such large $n$ correction terms, one could argue that the essential results of that approximation would be correct. But the sum of (3.9) runs over all $n$; and hence this simplifying approximation is clearly incorrect.

In fact, what this does suggest is that the model needs to be refined so that a new set of functions $H_n(mb)$, whose asymptotic $n$ dependence is sufficiently weak, should replace the $K_n(mb)$ of (3.9), in order to permit the interchange of limits stated abvove. This can be accomplished if one imagines that the neglected self--energy structure along each $\overline\Delta_c[\pi]$ line is included, with a net effect of damping away large $\tau$ contributions to the previous representations. It is at this point that one realizes that the neglected self--linkages along each $\overline\Delta_c$ must play an important role if non--ladder--graph processes and their non--leading order contributions are to be included. Such self--linkages are not necessary for the LL ladders, where a  `` nesting '' of the $n$ longitudinal momenta exchanged ( in order $G^{2n}$ ) between neighboring NVMs generates a factor of $1/n!$, so that the sum over $n$ is finite ( and exponentiates ). But if one asks for the sum of non leading contributions of a fixed order, e.g., proportional to $\left(\ln(s/m^2)\right)^2/2!$, the number of terms which enters into the corresponding sum grows so rapidly with $n$, that it is difficult to believe that the sum will converge. In our calculation, which embraces the sum of all ladder and crossed rung graphs of every order, large $n$ convergence is associated with large $\tau$ convergence; and, for this, one is thus naturally directed to attempt to include previously neglected self--linkages.

The model used above for extracting the non perturbative forms of high momentum linkages between different $\overline\Delta_c$ is not necessarily the one appropriate for the less energetic self--linkages along each $\overline\Delta_c$; but were it used for the latter, an extra factor of $\exp[-iG^2\tau^2I/2]$, $I=\!\int_0^1\! \!d\lambda\int_0^1\! \!d\lambda'\, D_c(\xi^{(-)}(\lambda)-\xi^{(-)}(\lambda'))$ would appear in each $\tau$ integral, and for $Re\, I \not =  0$ and/or $Im\, I < 0$, would generate significant damping for very large $\tau$.

A  completely different, soluble model, is one in which the original, Fradkin variable statement \cite{eleven} of all possible self--linkages is exactly expressed by the exponential of $\displaystyle i{G^2/ m^2}\!\int_0^s\!ds_1\!\int_0^{s_1}\!ds_2\, D_c\left(\int_{s_1}^ {s_2}\!ds'\,v(s')\right)$, and where the latter quantity is then approximated in a `` no recoil '' fashion by $\displaystyle i{G^2/ m^2}\!\int_0^s\!ds_1\!\int_0^{s_1}\!ds_2\, D_c\left(v_0(s_1 - s_2)\right)$, with $v_0$ corresponding to an `` averaged '' NVM 4--velocity, such that $v_0^2 = -1$. For zero mass propagator, it is well known that this propagator can be expressed exactly by 
$$
D_c(z) = \left({i\over 4\pi^2}\right){1\over z^2+i\varepsilon}\,\biggl\vert_{\varepsilon\to 0^{+}}\ ,\ \ \ D_c(v_0s_{12})\to-\left({i\over 4\pi^2}\right){1\over (s_1-s_2-i\overline\varepsilon)^2}\,\biggl\vert_{\overline\varepsilon\to 0^{+}}
$$
and, as evaluated elsewhere \cite{twelve}, the corresponding, self--linkage computation for linkages by a scalar field $\pi(x)$ yields the factors $$e\,^{\displaystyle is\Lambda^2 \alpha_G/\pi}\, e\,^{\displaystyle -{\alpha_G\over\pi}\ln(\Lambda^2/m^2)}(sm^2)^{-\displaystyle\alpha_G/\pi}$$
with momentum cut--off $\Lambda^2=(\overline\varepsilon)^{-1}$.

In sequence, these terms correspond to a model--dependent mass renormalization, a wave--function renormalization, and a damping of the $s$--integrand for large $s$. It is this latter factor which is of interest here, which damping remains after the $s\to-i\tau$ variable change introduced above. The model is not particularly realistic; but it again displays damping at large $s$, or $\tau$ values.

Let us therefore assume that, in general, such large $\tau$ damping does result from previously neglected self--linkages along each line; and take the simplifying step of inserting an effective, upper cut--off $q/m^2$ in the $\tau$ integral of (3.8), corresponding to the largest value of $\tau$ that enters when self--linkages are included
\be
\displaystyle\int_0^{q/m^2}{d\tau\over \tau}\,\tau^n\,e\,^{\displaystyle -m^2\tau - b^2/4\tau}\equiv 2\left({b\over 2m}\right)^nH_n(mb)
\ee
In effect, $K_n(mb)$ will then be replaced by $H_n(mb)$, a real, positive quantity with an upper bound given by
\be
H_n(mb) < {1\over  2}\left({2q\over 2mb}\right)^n\int_0^{q}{dt\over t}\,\,e\,^{\displaystyle -t - (mb)^2/4t}
\ee
What is the dimensionless quantity $q$? It can depend on the relevant, renormalized parameters of the theory, an effective $\alpha_G$, $m$, and $b$. If $q$ is chosen as a constant, $q_0$, then the summation of (3.9) will yield an approximate factor of $s\,^{\displaystyle(q_0^2\alpha_G/8\pi(mb)^2)}\overline K_0^2(mb)$ where $\overline K_0$ differs from $K_0$ in that its defining integral has an upper limit $q_0$, rather than $\infty$. But it is easy to show that, for $mb\gg 1$, there is ( exponentially ) little difference between $\overline K_0$ and $K_0$, so that an argument similar to that given in Appendix A produces the quantity $ {\displaystyle(q_0^2\alpha_G/8\pi(mb_0)^2)}\ln(s/m^2)\sim 2mb_0$, so that $b_0\sim \left[\ln(s/m^2)\right]^{1/3}$, and $\sigma_{\scriptscriptstyle TOT}\sim \left[\ln(s/m^2)\right]^{2/3}$. If, however, $q$ is assumed to grow linearly with $mb$,  $q\sim mb$, then the same argument reproduces the old Cheng--Wu result.

It seems that whichever form one adopts, for any $q\sim (mb)^i$, with $0<i<1$, one will find the tower--graph prediction of a slow--rising $\sigma_{\scriptscriptstyle TOT}$. Hence, the inclusion of all crossed -- as well as ladder-- graphs, in this model version of the tower graphs which requires strong, proper--time damping attribuable to the self--linkage graphs, generates a slowly rising $\sigma_{\scriptscriptstyle TOT}$; and for one special choice of cut--off, $q\sim mb$, it reproduces the form of the original Cheng--Wu result. This model is surely crude -- and was so from the beginning -- but crudeness does not necessary preclude correctness; and the predictions of the next Section could, conceivably, lead to qualitatively correct Physiccs.

For $mb\ll 1$, the transvers cut--off $\gamma$ should be taken as the inverse of an appropriate mass, and not as the smaller impact parameter; that is, $\gamma$ should always be chosen as the largest, relevant quantity with the dimension of length. In this region, the corresponding sum of the tower--graph contributions to the eikonal of (3.9) does not appear to converge, since the leading term of $K_n(mb)$ is $(1/2)(n-1)!/(mb/2)^n$ for small $mb$, a situation unchanged by the replacement of $K_n$ by $H_n$ if the ( large ) self--linkage cut--off $q$ can no longer be proportional to $(mb)^i$. What this indicates is that this eikonal is totally absorptive at small $b$; and just as in elementary, potential--theory calculations, it is to be replaced by a sufficiently large number, $\eta$, such that the entire contribution to 
\be
\Delta_2\sigma_{\scriptscriptstyle TOT} = \left({4\pi\over m^2}\right)\int_0^1\!dx\,x\left[1-e\,^{\displaystyle -\eta(s,x)}\right]
\ee
is just the `` black disk '' amount: $\Delta_2\sigma_{\scriptscriptstyle TOT} = 2\pi/m^2$. In this way, the complete tower--graph contributions  for ladder and crossed rung graphs again produce a slowly rising $\sigma_{\scriptscriptstyle TOT}$ with a maximum growth given by the old Cheng--Wu result. In the following Section, we choose for simplicity $q = mb$, so that $K_n\to H_n\to H_{n_{max}}\to K_0$; but quite similar results will follow for any other choice of $q\sim (mb)^i,\ 0<i<1$.
\vskip1cm   
{\bf\section{Summing all the Eikonal Graphs}}
\setcounter{equation}{0}
We now turn to the second, unanswered question of the eikonal approach to high--energy scattering: what is the effect, as posed in this model, of summing over all the remaining $Q_n,\ n>2$ ?

It is certainly possible to calculate each of the remaining $Q_n$ from their definition, as in (2.7), but enforcing the requirement of `` connectedness '' becomes tedious. It is much simpler to expand the  right hand side of (2.6) in power of $g^2$, to perform the needed functional operations on the $n^{th}$ term of that expansion, and then -- if possible -- to sum the results. One has
\beq
&e\,^{\displaystyle i\chi} = \displaystyle \sum_{n=0}^{\infty}{1\over n!}(ig^2)^n\,e\,^{\displaystyle-i\sum_{a>b}\int \!\!{\delta\over\delta\pi_a}D_c{\delta\over\delta\pi_b}}\nonumber\\
&\displaystyle\times\biggl(\int \!\!f_I\,{\overline\Delta_c}[\pi_a]\,f_{II}\biggr)\cdots\biggl(\int \!\!f_I\,{\overline\Delta_c}[\pi_n]\,f_{II}\biggr)\biggl\vert_{\pi_l\rightarrow 0}
\eeq
with the multiple factors of $( \displaystyle\int \!\!f_I\,{\overline\Delta_c}\,f_{II})$ occurring a total of $n$ times; (4.1) is the form that this $n^{th}$ functional approximation takes when all radiative corrections along each NVM are suppressed. The latter will, of course, be necessary, and will be introduced as needed.

For clarity, we work out the $n=3$ term, state the form of the $n=4$ term, and then infer the general result. Using the notation of the previous Section, this is
\beq
{}&\displaystyle{1\over 3!} \biggl(i{g^2(p_1\!\cdot\!p_2)\over 16\pi^2}\biggr)^3\int_0^{\infty}{ds_a\over s_a^2}\int_0^{\infty}{ds_b\over s_b^2}\int_0^{\infty}{ds_c\over s_c^2}\,e\,^{\displaystyle -i(s_a+s_b+s_c)m^2}\nonumber\\
&\hskip-2truecm\displaystyle\times\int\!\!\!\int_{-\infty}^{+\infty}\!\!d\bar s_a\,d\bar t_a\int\!\!\!\int_{-\infty}^{+\infty}\!\!d\bar s_b\,d\bar t_b\int\!\!\!\int_{-\infty}^{+\infty}\!\!d\bar s_c\,d\bar t_c\\
&\hskip-1truecm\times\displaystyle \exp\biggl[{\displaystyle {i\over 4s_a}(z_{12}-\bar s_ap_1+\bar t_ap_2)^2+ {i\over 4s_b}(z_{12}-\bar s_bp_1+\bar t_bp_2)^2+{i\over 4s_c}(z_{12}-\bar s_cp_1+\bar t_cp_2)^2}\biggr] \nonumber\\
&\hskip-3truecm\displaystyle\times\, \exp\biggl[-i \sum_{a>b}\int\!{\delta\over\delta\pi_a}D_c{\delta\over\delta\pi_b}\biggr]\nonumber\\
&\displaystyle\times \exp\biggl[ -iGs_a\!\int_0^1\! \!d\lambda_a\, \pi_a(\xi_a^{(-)}(\lambda_a)) -iGs_b\!\int_0^1\! \!d\lambda_b\, \pi_b(\xi_b^{(-)}(\lambda_b))-iGs_c\!\int_0^1\! \!d\lambda_c\, \pi_c(\xi_c^{(-)}(\lambda_c))\biggr]\biggl\vert_{\pi_{l}\to 0}\nonumber
\eeq
As before, each linkage operator -- and there are here $n(n-1)/2=3$ of them~-- generates a term of form 
$$
\exp\biggl[-\alpha_G\Bigl({m^2s_as_b\over 4\pi\gamma^2}\Bigr)\ln(s/m^2)\biggr]
$$
-- the result of (3.6), but without the factor of $-1$ -- while the $\int d\overline s_a\cdots \int d\overline t_c$ integrations remove all non transverse $(z_{12})^2$ dependence, generating
$$ \Bigl({8\pi\over s}\Bigr)^3(s_as_bs_c)\,\,e\,^{\displaystyle ib^2/4\left({1\over s_a}+{1\over s_b}+{1\over s_c}\right)}
$$
so that this $n=3$ contribution to $\exp(i\chi)$ produces
\beq
&\displaystyle{(-i\alpha_g)^3\over 3!}\int\!\!\!\int\!\!\!\int_0^{\infty}{ds_a\over s_a}{ds_b\over s_b}{ds_c\over s_c}\,e\,^{\displaystyle -i(s_a+s_b+s_c)m^2+i{b^2\over 4}\left({1\over s_a}+{1\over s_b}+{1\over s_c}\right)}\nonumber\\
&\times\displaystyle\exp\biggl[-{m^2\alpha_G\over 4\pi\gamma^2}\Bigl(s_as_b+s_bs_c+s_cs_a\Bigr)\ln(s/m^2)\biggr]\nonumber
\eeq
Expanding each of the $n(n-1)/2$ factors of form  $\displaystyle\exp\Bigl[-\Bigl({m^2\alpha_G\over 4\pi\gamma^2}\Bigr)s_as_b\ln(s/m^2)\Bigr]$, one obtains
\be
\hskip-0.5truecm\displaystyle{(-2i\alpha_g)^3\over 3!}\!\!\sum_{n_{1,2,3}=0}^{\infty}\!\!{{\Bigl[\displaystyle{\alpha_G\over 8\pi}{b^2\over\gamma^2}\ln(s/m^2)\Bigr]\over n_1!\,n_2!\,n_3!}}^{n_1+n_2+n_3}\hskip-1.5truecm  K_{n_1+n_3}(mb)K_{n_1+n_2}(mb)K_{n_2+n_3}(mb)
\ee
In contrast, for $n=4$, one would find  $n(n-1)/2=6$ summations over $n_1,\cdots,n_6$:
\be
\hskip-0.5truecm\displaystyle{(-2i\alpha_g)^4\over 4!}\hskip-0.5truecm\sum_{n_1,\cdots,n_6=0}^{\infty}\!\!{{\Bigl[\displaystyle{\alpha_G\over 8\pi}{b^2\over\gamma^2}\ln(s/m^2)\Bigr]\over n_1!\,\cdots\,n_6!}}^{n_1+\cdots+n_6}\hskip-1.3truecm  K_{N_1}(mb)K_{N_2}(mb)K_{N_3}(mb)K_{N_4}(mb)
\ee
where $\displaystyle N_i=\sum_{j=1}^6n_j-n_i$

Can one sum the series of these terms ? Again, there is a great simplification for $\gamma=b$, $mb>1$, and ( at this point, introducing the self--linkages along each $\overline\Delta_c[\pi]$, which effectively introduce a $q/m^2$ cut--off into each proper--time integeral, so that, for the simplest case where $q=mb$, each ) $K_n\to(\pi/2mb)^{-1/2}\exp(-mb)$. In this way, (4.3) becomes
$$
 \displaystyle{(-2i\alpha_g)^3\over 3!}\Bigl({\pi\over 2mb}\Bigr)^{3/2}e\,^{\displaystyle -3mb}\,(s/m^2)^{\displaystyle3\alpha_G/8\pi}
$$
while (4.4) yields
$$
 \displaystyle{(-2i\alpha_g)^4\over 4!}\Bigl({\pi\over 2mb}\Bigr)^{4/2}e\,^{\displaystyle -4mb}\,(s/m^2)^{\displaystyle6\alpha_G/8\pi}
$$
and the general pattern is clear:
\beq
&\displaystyle e\,^{\displaystyle i\chi}\Bigl\vert_{mb>1}\simeq 1 + i\chi_1\vert_{mb>1}\\
&\displaystyle  + \sum_{n=2}^{\infty}{(-2i\alpha_g)^n\over n!}\Bigl({\pi\over 2mb}\Bigr)^{n/2}e\,^{\displaystyle -nmb}\,(s/m^2)^{\displaystyle({n(n-1)\over 2})(\alpha_G/8\pi)}\nonumber
\eeq

The next task is to give a meaningful value to the unusual sum of (4.5). Because of the factor $(-i)^n$, there will be an alternation in the signs of the terms that comprise $Re\,\exp[i\chi]$; and because the $s$--dependence has the form $s^{an(n-1)/2}$, these alternations can be important. To see this in detail, we make use of a representation \cite{one} expressed in terms of a convergent power series:
$$
\displaystyle  \sum_{n=2}^{\infty}{(-iZ)\over n!}^n\equiv F(Z) = e\,^{\displaystyle -iZ}-1+iZ\ ,\ \ \ \ \ Z={\overline x\over y}\,e\,^{\displaystyle 2\alpha\beta}
$$
and note that
$$
\displaystyle S(\overline x,y)\equiv{1\over\sqrt\pi}\int_{-\infty}^{\infty}\!\!d\alpha \,e\,^{\displaystyle-\alpha^2}F\Bigl({\overline x\over y}\,e\,^{\displaystyle 2\alpha\beta}\Bigr)= \sum_{n=2}^{\infty}\Bigl({-i\overline x\over y}\Bigr)^n{1\over n!}\,e\,^{\displaystyle n^2\beta^2}
$$
This is the same series as that of (4.5), if the identifications
$$\displaystyle y=\Bigl({s\over m^2}\Bigr)^{\alpha_G/16\pi}\ ,\ \ \ \ \ \beta^2=\ln y\ ,\ \ \ \ \ \overline x=\alpha_g\sqrt{2\pi\over mb}\,e\,^{\displaystyle -mb}
$$
are made, and we may therefore replace (4.5) by the integral
\be
\hskip-0.3truecm\displaystyle e\,^{\displaystyle i\chi}\Bigl\vert_{mb>1}\to 1 + i\chi_1+ {1\over\sqrt\pi}\int_{-\infty}^{\infty}\!\!d\alpha \,e\,^{\displaystyle-\alpha^2}\biggl\{e\,^{\displaystyle -i\left({\overline x\over y}\,e\,^{\displaystyle 2\alpha\beta}\right)}-1+i{\overline x\over y}\,e\,^{\displaystyle 2\alpha\beta}\biggr\}
\ee
so that the contribution of (4.6) to $\sigma_{\scriptscriptstyle TOT}$ is given by
\be
\displaystyle\Delta_1\sigma_{\scriptscriptstyle TOT}={2\over\sqrt\pi}\int_{1/m}^{\infty}\!\!d^2b\int_{-\infty}^{\infty}\!\!d\alpha \,e\,^{\displaystyle-\alpha^2}\biggl[1-\cos\left({\overline x\over y}\,e\,^{\displaystyle 2\alpha\beta}\right)\biggr]
\ee
It will be convenient to make the variable change: $\alpha\to\displaystyle{1\over 2\beta}\ln\left({yz\over\overline x}\right)$. Approximating $\ln(\overline x)$ by $-mb$, and with the additional variable change : $mb=\beta^2(u-1)$, and with $x = 4\beta^2$, one finds
\be
\hskip-0.3truecm\displaystyle\Delta_1\sigma_{\scriptscriptstyle TOT}={\sqrt\pi\over 4m^2}\,x^{3/2}\int_{1+4/x}^{\infty}\!\!du\,(u-1)\,e\,^{\displaystyle-xu^2/16}\int_{0}^{\infty}\!\!dz\, {(1-\cos z)\over z^{1+u/2}}\,e\,^{\displaystyle -{1\over x}\ln ^2z}
\ee
where $\Delta_1\sigma_{\scriptscriptstyle TOT}$ is again that contribution to $\sigma_{\scriptscriptstyle TOT}$ arising from $b>1/m$. These integrals are real and positive definite, but cannot be evaluated analytically, so that a numerical approach is necessary. However, it is fairly easy to approximate (4.8), so that the evaluation can be done in an analytic manner, and for this, one first rewrites the $z$ integral of (4.8) as $\displaystyle\int_{0}^{\infty}\!\!dz\,\exp[-f(z)][1-\cos z]$, where $\displaystyle f(z) = [ {1\over x}\ln ^2z + (1+{u\over 2})\ln z]$, and where a typical graph of $\exp[-f(z)]$, for $u=x=2$, is displayed in Fig.4.

\begin{figure}[htpb]
\centering\includegraphics[height=6cm,clip=true]{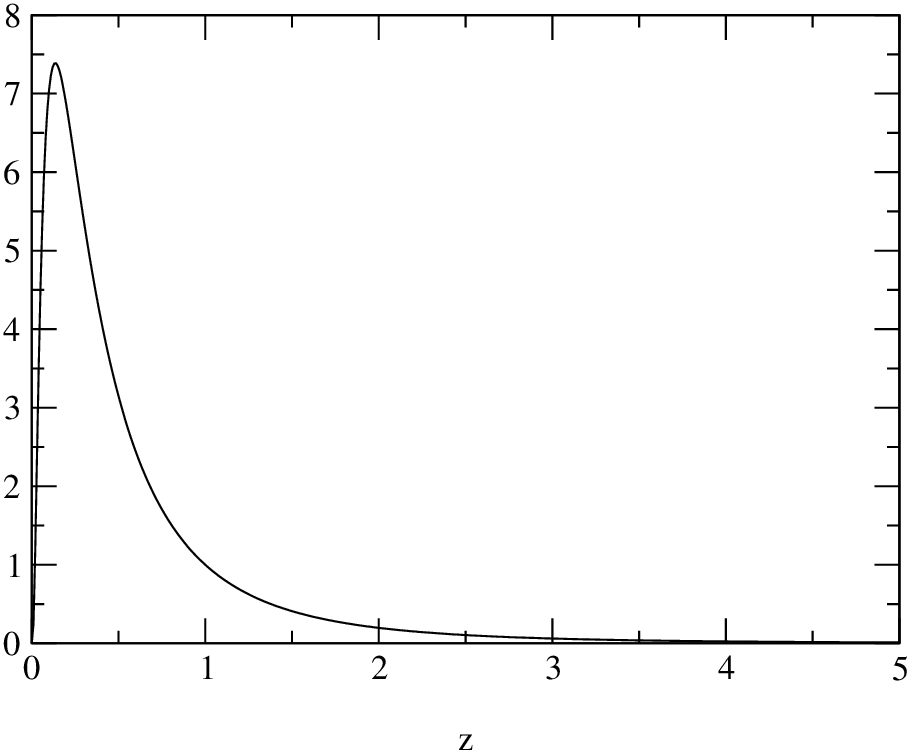}
\caption{A plot of $\displaystyle \exp\,\{ -[ {1\over x}\ln ^2z + (1+{u\over 2})\ln z]\}$ vs. z for $x=u=2$}
\end{figure}

 The peak of that curve occurs at a value $z_0<1$, and its peak height is $\exp[-f(z_0)]$. For $z>z_0$, the curve has a rough, Gaussian appearance, but for $z$ values less than $z_0$ this is not the case, for the curve vanishes rapidly as $z\to 0$. However, $\exp[-f(z)]$ is multiplied by $2\sin^2(z/2)$, which vanishes as $z\to 0$; and therefore it is not too inaccurate to replace $\exp[-f(z)]$ by a simpler Gaussian form about $z_0$, since the contributions for $z<z_0$ are going to be very small; and we will therefore write $f(z)\simeq f(z_0) + (1/2)(z - z_0)^2f''(z_0)$, with $z_0$ determined by the condition $f'(z_0) = 0 : z_0 = \exp[-(x/2)(1+u/2)]$. The argument of this exponential factor is, following from the lower limit of the $u$ integral, always more negative than $-(1+x/2)$, so that $z_0$ is always small, especially for large values of $x$. If no further approximation is used, the $\int \!dz$ would be given in terms of probability integrals, $\Phi (x)$.

But the smallness of $z_0$, especially in a region where the vanishing of $(1 - \cos z)$ removes most of the error, now suggests an additional approximation~: replace $\displaystyle\int_{0}^{\infty}\!\!dz$ by $\displaystyle{1\over 2}\int_{-\infty}^{\infty}\!\!dz$, so that the integral can be evaluated immediately as
\be
\displaystyle {1\over 2}\,e\,^{\displaystyle-f(z_0)}\sqrt{2\pi\over f''(z_0)}\biggl[1 - e\,^{\displaystyle-{1\over 2f''(z_0)}}\cos z_0\biggr]
\ee
which generates a convergent integrand for large $u$. Further, since $z_0<1$, expansion of the $\cos z_0$ of (4.9) will give corrections to the result obtained by setting $z_0\to 0$, which are smaller by exponential dependence on $x$; and hence it is legitimate to replace $\cos z_0$ in (4.9) by unity.

Although the resulting integrand is reminiscent of that of (2.5), the techniques used for the approximate evaluation of the latter will not work here; rather, because $u_{min} = 1 + 4/x$, and $x>0$, $h(u,x) = (x/4)\exp[-x(1 + u/2)]$ is $<1$ for any value of $x$, and the terms of (4.9) may be expanded in powers of $h(u,x)$, with the linear term in $h$ generating
\be
\displaystyle\Delta_1\sigma_{\scriptscriptstyle TOT(s)}\simeq \biggl({2\pi\over m^2e^2}\biggr)\,x\,e\,^{\displaystyle-{7x\over 4}}\ ,\ \ \ \ \ \ \ x = \Bigl({\alpha_G\over 4\pi}\Bigr)\ln\Bigl({s\over m^2}\Bigr)
\ee
and with corrections which are smaller by exponential factors of $x$.

The graph of (4.10) in Fig.5 tells the entire story. As $x$ increases from zero, $\Delta_1\sigma_{\scriptscriptstyle TOT}$ rises linearly with $\ln(s)$, peaks at $x = 4/7$, and then falls off exponentially. There are therefore, as suspected, sufficient cancellations within this model to remove the rising total cross section of the tower graphs. The details of the calculation are certainly subject to correction; but the result of (4.10) does suggest serious cancellations away from the tower graph result. If, for example, one imagines that scattering at $s\sim 1\, TeV^2$ corresponds to values of $x\sim .3$, then the peak of $\Delta_1\sigma_{\scriptscriptstyle TOT}$should appear at $x\sim .6$, corresponding to a doubling of the value of $\ln s$. Such an energy is probably higher than that of the so--called `` cosmic ray point '', and if so there is little poss
ibility of its ever being measured directly.

\begin{figure}[htpb]
\centering\includegraphics[height=6cm,clip=true]{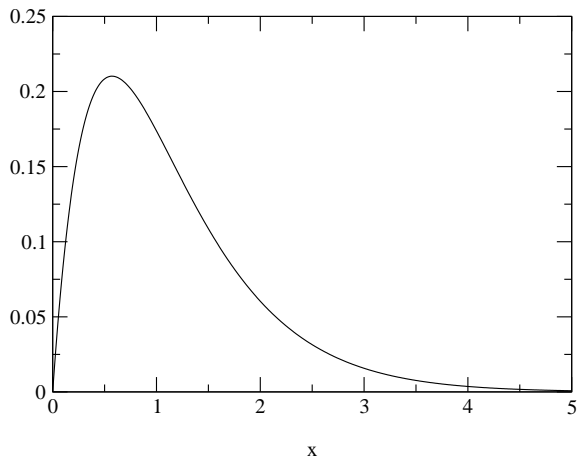}
\caption{A plot of $[(me)^2/2\pi]\sigma_{\scriptscriptstyle TOT}(s)$ vs. $x=(\alpha_G/4\pi)\ln(s/m^2)$}
\end{figure}

Without any further calculation, one expects the $mb<1$ contributions to reduce ( as for the tower graphs, but more quickly ) to the same value of  $\Delta_2\sigma_{\scriptscriptstyle TOT} = 2\pi/m^2$, which result has no real bearing on the question of $\sigma_{\scriptscriptstyle TOT} $ at asymptotic energies.

Could this model possibly apply to QCD ? Not directly, for scalar pions are neither gluons nor closed quark--loops. But if one's imagination is allowed free rein, and if scalar pions are repleceable by gluons ( as the basic elements of gluonic jets ), then one might imagine that timelike renormalization effects of the asymptotically free QCD could decrease $\alpha_G$ by a factor of $\ln(s/m^2)$, so that $x = (\alpha_G/4 \pi)\ln(s/m^2)$ can never increase significantly past a constant amount, say $x_{max}$; and hence that $\sigma_{\scriptscriptstyle TOT}$ would become a non zero constant, larger than $2\pi/m^2$, at $x_{max}$, and would stay at that value for larger values of $s$. Of course, this is rampant speculation, and one does not yet know the answer for real QCD. Modulo questions of mathematical rigor, the arguments of this Section suggests that, at truly asymptotic energies, and if no further fields with higher mass quanta appear, total cross sections could become constants, either large constants if timelike asymptotic freedom holds, or smaller constants if it does not; but they need not continue to grow in the form of the Froissart bound.
\vskip2cm
{\bf Acknowledgments}
\vskip1cm
It is a pleasure to thank Profs. A. Thomas and T. Williams for their kind hospitality, and O. Leitner for technical (electronic) assistance, at CSSM, Adelaide University SA, where this work was completed.

\vskip1cm
\def\theequation{A.\arabic{equation}}
{\bf\section*{Appendix A:  The Cheng--Wu eikonal}}
\setcounter{equation}{0}
\appendix
It will be useful to illustrate how the ladder--graph tower--eikonals generates their $\sigma_{\scriptscriptstyle TOT}(s)$, and for this we shall consider the simplest case of all, that of the original Cheng--Wu eikonal, for which one may write $i\chi = -\rho(s,b)$, $\rho(s,b) = a\,s^{\alpha}\exp(-\mu b)$. Eq. (2.3) may then be written as 
\be
\sigma_{\scriptscriptstyle TOT}(s) = 2\, \!\int \!d^2b\,\Bigl[1 - e\,^{\displaystyle -\rho(s,b)}\Bigr]
\ee
and to evaluate the form of the resulting $s$ dependence it is convenient to define a quantity $b_0(s)$ by the relation: $1 = \rho(s,b_0(s))$; that is, $b_0(s)\simeq (\alpha/\mu)\ln\left(s/m^2\right)$ is that value of impact parameter where any increase of $\rho$ with increasing $s$ is just counterbalanced by the damping with respect to $b$. One sees from (A.1) that for $b<b_0$, $\rho$ is large and the $\exp(-\rho)$ is small, so that this contribution gives essentially 
\be
 2\, \!\int_0^{b_0} \!d^2b\,[1] \simeq 2\pi \,b_0^2(s) \sim \ln^2(s/m^2) + \cdots
\ee
In contrast, for $b>b_0$, $\rho$ is small, and the exponential of (A.1) may be expanded, and that portion of the integral approximated by
\be
 2\, \!\int_0^{\infty} \!d^2b\,\rho  \sim \ln(s/m^2) + \cdots
\ee
which is down by one factor of $\ln(s)$ compared to the leading $s$ dependence of $\sigma_{\scriptscriptstyle TOT}$, which arises from large $b\sim\ln(s)$.
\vskip2truecm
\def\theequation{B.\arabic{equation}}
{\bf\section*{Appendix B: The `` scalar '' laser solution for ${\bf G_c[A]}$}}
\setcounter{equation}{0}
\appendix
The word `` laser '' used in this discussion is really a misnomer, for it should properly be replaced by `` electromagnetic plane wave '' ( epw~); but we ask the reader's indulgence for this simple idealization, which is reasonable as long as the perpendicular dimensions of the laser beams under question are much larger than the dimensions of the charged particle on which they are acting, or on the transverse distances over which the particle is to move.

The simplest, idealized, plane wave `` laser '' solution occurs for the scalar $G_c[A]$ with scalar interaction $A(x)\to A(k\!\cdot\!x)$, where the functional form of $A$ is arbitrary, but $k^2 = \vec k^2 - k_0^2 = 0$; although less complicated than the full laser solution of QED, the essential features of solubility are the same.

Here, one uses the functional, Fradkin solution
\beq
&\displaystyle G_c(x,y\vert A) = i\!\int_0^{\infty}\!\!ds\,e^{\displaystyle -ism^2}\,e^{\displaystyle -i\!\int_0^s\!ds'{\delta^2\over \delta v_{\mu}^2(s')}}\nonumber\\
&\\
\noalign{\vskip-0.5cm}
&\displaystyle\times\, \delta(x-y+\!\int_0^s\!ds'v(s'))\,e^{\displaystyle -ig\!\int_0^s\!ds'A(y-\int_0^{s'} \!\!ds''v(s''))}\nonumber
\eeq
and inserts an expression for unity under the integrals of (B.1) with the express purpose of extracting the $v$ dependence that appears in the argument of $A$,
\beq
&\displaystyle G_c(x,y\vert A) = i\!\int_0^{\infty}\!\!ds\,e^{\displaystyle -ism^2}\!\!\int\!{d^4p\over (2\pi)^4}\,e\,^{\displaystyle ip\!\cdot\!(x-y)} N'\!\!\int\!\! d[u]\!\int\!\! d[\Omega]\,e\,^{\displaystyle i\!\!\int_0^s\!\!u\Omega}{\cal F}[u]\nonumber\\
&\\
\noalign{\vskip-0.5cm}
&\displaystyle\times e^{\displaystyle -i\!\int_0^s\!{\delta^2\over \delta v^2}}e\,^{\displaystyle i\!\int_0^s\!\!ds'v_{\mu}(s')[\,p_{\mu}-\int_{s'}^s\!\!ds''\,k_{\mu}\Omega(s'')]}\bigg\vert_{v\to 0}\nonumber
\eeq
where $\displaystyle{\cal F}[u] = \exp\Bigl[-ig\!\!\int_0^s\!\!ds'A((k\!\cdot\!y-u(s'))\Bigr]$, and Abel's replacement of $\displaystyle\int_0^s\!\!ds'\,\Omega(s')\!\int_0^{s'}\!\!ds''v_{\mu}(s'')$ by $\displaystyle\int_0^s\!\!ds'\,v_{\mu}(s')\!\int_{s'}^{s}\!\!ds''\,\Omega(s'')$ has been used. The Fradkin functional operation of the second line of (B.2) is now immediate, and yields
\be
\hskip-0.3truecm\exp\biggl[- i\!\int_0^s\!\!ds'\Bigl[\,p_{\mu}-k_{\mu}\!\int_{s'}^s\!\!ds''\,\Omega(s'')\Bigr]^2\biggr] = \exp\biggl[- isp^2+2ip\cdot \!k\!\int_0^s\!\!ds's'\,\Omega(s')\biggr]
\ee
where the inverse of Abel's trick has been used. The essential feature which guarantees solubility is then apparent: because $k^2 = 0$, there is no quadratic $\Omega$ dependence, and its functional integral yields the simple delta functional: $\delta\left[u(s')+2s'p\!\cdot \!k\right]$. Then, $\int\! d[u]$ is immediate, replacing ${\cal F}[u]$ by ${\cal F}[-2s'p\!\cdot \!k]$,
\beq
&\displaystyle G_c(x,y\vert A) = i\!\int_0^{\infty}\!\!ds\,e^{\displaystyle -ism^2}\!\!\int\!{d^4p\over (2\pi)^4}\,e\,^{\displaystyle ip\!\cdot\!(x-y)-isp^2} \nonumber\\
&\\
\noalign{\vskip-0.5cm}
&\displaystyle\times e^{\displaystyle -ig\!\int_0^s\!ds'A(k\!\cdot \!y+2s'p\!\cdot \!k)}\nonumber
\eeq
Eq. (B.4) can be further simplified. In essence, with $z=x-y$, one requires the integral
\be
\int\!{d^4p\over (2\pi)^4}\,e\,^{\displaystyle ip\!\cdot\!z-isp^2}Q(p\!\cdot \!k)
\ee
where $Q$ may be read off directly from (B.4). One proceeds in a manner analogous to that used above by introducing under the integrals of (B.5) a factor of unity,
$$
1=\int_{-\infty}^{\infty}\!\!du\int_{-\infty}^{\infty}{d\omega\over 2\pi}\,e\,^{\displaystyle i\omega(u-k\!\cdot\!p)}
$$
from which one obtains
\beq
&\displaystyle G_c(x,y\vert A) = i\!\int_0^{\infty}\!\!ds\,e^{\displaystyle -ism^2}\int_{-\infty}^{\infty}\!\!du\,e^{\displaystyle -ig\!\int_0^s\!ds'A(k\!\cdot \!y+2s'u)}\int_{-\infty}^{\infty}{d\omega\over 2\pi}\,e\,^{\displaystyle i\omega u}\nonumber\\
&\\
\noalign{\vskip-0.5cm}
&\displaystyle\times \!\!\int\!{d^4p\over (2\pi)^4}\,e\,^{\displaystyle -isp^2+ip\!\cdot\!(z-k\omega)} \nonumber
\eeq
The last line of (B.6) is a simple Gaussian, yielding 
$$
\biggl({-i\over 16\pi^2s^2}\biggr)\,e\,^{\displaystyle i(z-k\omega)^2/4s}
$$
and one notes that because $k^2=0$, the $\omega^2$ term in the expansion of the exponential factor is missing, so that the $\omega$ integration generates $\delta(u-k\cdot \!z/2s)$ permitting the $u$ integral to be performed. With the variable change $\lambda = s'/s$, one obtains
\be
\displaystyle G_c(x,y\vert A) = {1\over 16\pi^2}\!\int_0^{\infty}\!{ds\over s^2}\,e^{\displaystyle -is[m^2+g\!\int_0^1\!d\lambda\,A(k\!\cdot \!\xi(\lambda))]+i(x-y)^2/4s}
\ee
where $\xi_{\mu}(\lambda)=\lambda x_{\mu}+(1-\lambda)\,y_{\mu}$ represents the straight--line path between the points $y_{\mu}$ and $x_{\mu}$. Finally, one realizes that (B.7) is just the ordinary, scalar, causal, `` free particle '' boson propagator $\Delta_c(x-y,m^2)$, but with its mass $m^2$ replaced by a position--dependent mass$^2$: $m^2\to M^2=m^2+g\!\int_0^1\!d\lambda\,A(k\!\cdot \!\xi(\lambda))$. If one now replaces $k_{\mu}$ by $k_{\mu}/\omega$, and $A(x)$ by $\pi(x)$, one obtains the model Green's function $\overline\Delta_c[\pi]$ used ion the text.






\begin{thebibliography}{**}
\bibitem{three} The original work was due to Cheng and Wu ( with references to their several papers appearing in \cite{eight} ), who first suggested the possibility of slowly--rising cross sections that satisfy the Froissart bound. Other calculations in other contexts soon followed, e.g., the papers quoted in \cite{nine}.
\end{thebibliography}
\end{document}